\documentstyle[floats,twocolumn,prl,aps]{revtex}
\begin{document}

\title{1/$\omega$-flux-noise and dynamical critical properties of 
two-dimensional  XY-models}

\author{P. H. E. Tiesinga$^{1}$\cite{*}, T. J. Hagenaars$^{1,3}$, J. E. van
Himbergen$^1$ and Jorge V. Jos\'{e}$^{2}$}

\address{{\it $^1$Instituut voor Theoretische Fysica,\\ Princetonplein 5,
Postbus 80006, 3508 TA Utrecht, The Netherlands\\
$^2$ Physics Department and Center for the Interdisciplinary
Research on Complex Systems \\ Northeastern University, Boston, MA
02115\\
 $^3$ Institut f\"ur Theoretische Physik,
Universit\"at W\"urzburg,\\Am Hubland, 97074 W\"urzburg, Germany 
}}

\author{\parbox{397pt}{\vglue 0.3cm \small
We have numerically studied the  dynamic correlation functions in 
thermodynamic equilibrium of two-dimensional $O(2)$-symmetry models 
with either bond (RSJ) or site (TDGL) dissipation as a function of 
temperature $T$. We find  that above the critical temperature the  
frequency  dependent flux noise  
$S_{\Phi}(\omega)\sim \vert 1+{(\omega/\Omega)}^2\vert^{-\alpha (T)/2}$, with
$0.85\leq \alpha (TDGL)(T)\leq 0.95$ and $1.17\leq \alpha (RSJ)(T) \leq 1.27$,
while the dynamic critical exponents  $z(TDGL)\sim 2.0$ and $z(RSJ)\sim 0.9$.
Contrary to expectation the TDGL results are in closer agreement 
with the experiments in 
Josephson-junction arrays by Shaw {\em et al.}, than those from
the RSJ model.   We find that these results 
are related to anomalous vortex diffusion through vortex {\it clusters}.
 }}

\maketitle
There is a general theoretical consensus that the equilibrium critical
properties of two-dimensional Abelian symmetry statistical mechanical models
are properly described by the Berezinskii-Kosterlitz-Thouless 
(BKT) theory \cite{KT1,JKKN}. The theory is based on the idea that 
at low temperatures there are thermally excited vortex-antivortex 
pairs that unbind just above the BKT critical temperature. 
Experimental tests have for the most part been indirect, in out of 
equilibrium conditions. Successful comparisons of the BKT predictions 
to experiment were initially carried out in superfluid helium 
\cite{reppy} and in superconducting films  \cite{Beasley}.  Later 
measurements on Josephson-junction arrays (JJA) yielded 
current-voltage characteristics \cite{Resnick}, and dynamic impedance 
\cite{martikt} results that also agree with the BKT theory. To explain 
the experiments, the static BKT scenario was heuristically extended  
to non-equilibrium  situations \cite{AHNS}. One of the basic tenets of 
the dynamical BKT extensions is that vortices move {\it diffusively} 
with binding and unbinding of vortex pairs under the action of
the  external drive. Further phenomenological modifications to the dynamic
extension of the BKT theory, that reduced the number of fitting parameters,
have led to better correspondence to some of the experimental  
results on superconductors \cite{minr}. 

A very recent flux noise SQUID experiment by Shaw {\em et al.}  \cite{shaw} 
on proximity effect JJA  provided a direct experimental test of 
the {\it equilibrium} BKT theory, at $T_{BKT}<T$\cite{cap}. They found 
the flux noise to be white at the lowest frequencies, and  
proportional to $1/{\omega}^{\alpha}$,  with  $\alpha \approx 1$,
at intermediate ones rather than the $1/\omega^2$ expected by the 
phenomenological theories \cite{AHNS,minr}. They also carried out a dynamical 
scaling analysis of the noise function for different temperatures,
that yielded a dynamical exponent of $z\sim 2$. 
An earlier, less extensive, JJA experiment by Lerch {\em et al.} 
\cite{martihelv} found a $1/\omega$ behavior for
the $T<T_{BKT}$ regime, without the white noise frequency region.
In this paper we concentrate on the $T_{BKT}<T$ regime. 
We show that by numerically calculating quantities 
{\it closely} related to those measured in \cite{shaw}
we do get good agreement with their $\alpha$ and $z$ results
for a site dissipation model defined below \cite{nos1}.

There are two models that have been considered to study the arrays. 
One is the resistively-shunted Josephson (RSJ) array model 
and the other is related to the time-dependent
Ginsburg-Landau (TDGL) model. The RSJ model is constructed from 
the elementary RSJ equations for  single Josephson junctions 
that form the  array units, plus Kirchhoff's current conservation 
conditions at each lattice site. The JJA-RSJ model has been 
successful in explaining, for example, the experimental giant 
Shapiro steps that arise when the JJA is driven by a dc+ac 
current \cite{rev1}. The TDGL model is an alternative  dynamical model 
that has been used to describe the general critical dynamical 
properties of the XY-model \cite{xydyn} and also some aspects of 
JJA  behavior, including its flux noise \cite{beck,olss}. An important 
theoretical question is then to ascertain which model is best to
describe the experimental noise results. To find out,
we have studied the flux noise and dynamical scaling properties for
both models. We do indeed find that the flux noise is anomalous in
both cases,  but the TDGL results appear to be
closer to the experimental ones. We find that the vortices do
not move independently but in a sea of vortex clusters that modify their
diffusive properties.

The equation of motion for both models can be written in the general
Langevin equation form,
\begin{equation}
{\partial_t \theta({\bf r},t)} = \sum _{{\bf r'}}[\,\, \Gamma ({\bf r},{\bf r'})
\frac{\partial H[\theta]}{\partial
\theta ({\bf r'},t)}+W({\bf r},{\bf r'},t)\,\,].
\label{general}
\end{equation}
Here $\theta({\bf r},t)$ is the phase of the order parameter of
the superconducting island at site ${\bf r}$; 
$H[\theta ]=E_J\cos (\theta({\bf r'},t)-\theta({\bf r},t))$ 
is the Hamiltonian bond energy, with $E_J$ the Josephson coupling,
and $W({\bf r},t;{\bf r'},t')$ the noise function. The RSJ model has
$\Gamma_{RSJ}({\bf r},{\bf r'})\equiv \Gamma G({\bf r},{\bf r'})$.
In our units $\Gamma=1$. $G({\bf r},{\bf r'})$, the two-dimensional 
inverse lattice Laplacian, arises from the fact that the currents 
in the array are conserved at each lattice site.  The dissipation 
here is present in the junctions between the superconducting islands.
The random noise function 
$W_{RSJ}({\bf r}, t;{\bf r'}, t')= \eta ({\bf r}, t; {\bf r'}, t')$ 
is defined at each bond in the lattice with Gaussian properties
 $<\eta ({\bf r},t ; {\bf r'}, t')>=0$ and $<\eta ({\bf r},t ;{\bf r}+
{\bf e}_i, t')\eta ({\bf r'}, t ;{\bf r'}+{\bf e}_j, t')>=2T \delta_{i,j}
\delta({\bf r}-{\bf r'})\delta (t-t')$,
where ${\bf e}_j$ is a unit vector along the {\it j}th direction,
and $T$ is the dimensionless temperature. 

The TDGL model is defined by taking $\Gamma_{TDGL}({\bf r},{\bf r'})=
\Gamma \delta({\bf r}- {\bf r'})$ ($\Gamma=1$), and 
$W_{TDGL}({\bf r},t;{\bf r'},t')=\eta ({\bf r},t)
\delta({\bf r}-{\bf r'})\delta(t-t')$; 
here the noise is also white and defined at each island,
with $<\eta ({\bf r},t)>=0$ and $<\eta ({\bf r},t)
\eta ({\bf r'},t')>=2T \delta({\bf r}-{\bf r'})\delta (t-t')$. 
This model has been used before to study the critical dynamics of the 
XY-model. This is in part because it is computationally 
less demanding than the RSJ model, that needs time consuming 
matrix inversions to evaluate G(${\bf r},{\bf r'})$. 
We find that the macroscopic thermodynamic properties, like the helicity 
modulus and the phase correlation functions described below,  
are essentially the same for both models. However, as we show here, other 
dynamical equilibrium properties are different  for the two models, 
and much more so the out-of-equilibrium quantities \cite{visco}.

An important aspect of the Shaw {\em et al.} study  is that the SQUID used  
for measurements was smaller than the array size. The SQUID  measures 
the average net number  of thermally nucleated and annihilated vortices 
that randomly enter  or exit its effective area. We do exactly the same 
in our calculations, by considering  regions of the 
lattice smaller in  size than that of the full array simulated 
(see Fig.\ref{1}(b)). 
%%%%%%%%%%%%%%%%%%%%%%%%%%%%%%%
%{figure 1}
%%%%%%%%%%%%%%%%%%%%%%%%%%%%%%%%%%%
The space-time local vorticity at plaquette
${\bf R}$ is defined by $ 2\pi n({\bf R},t)=\sum_{{\cal P}({\bf R})}
\big[\theta({\bf r},t)-\theta({\bf r'},t)\big] \mbox{mod}(2\pi).$
Here ${\cal P}({\bf R})$ denotes an anti-clockwise sum around
plaquette ${\bf R}$. We imposed periodic boundary conditions (PBC)
in both directions in our equilibrium calculations. The average vorticity 
for the {\it i}th SQUID of area $\ell \times \ell$ is defined as
$N_\ell ^i(t)=\sum_{{{\bf R}\epsilon} (\ell \times \ell)_i} n({\bf R},t)$.
To get the net flux noise we average over all the $i$-SQUIDS  
that can be fitted in the array of size $L\times L$ 
(See Fig. \ref{1}(d).) The flux noise produced by the vortices in the 
time domain is then generally defined by
\begin{equation}
g_\ell^V(t)\equiv \frac{1}{N_{\ell}} \sum_i[\,\, \langle 
N_\ell ^i(t) N_\ell ^i(0) \rangle-
\langle
 N_\ell ^i(0) \rangle^2\,\, ].
\label{gvp}
\end{equation}
Here $N_\ell$ is the total number of $i$-SQUIDS of size $\ell\times\ell$, 
and the averages $<,>$ are carried out over the probability distribution for
the noise or over time. Based on the ergodic theorem, both averaging 
methods should and do give the same results, but the latter is easier 
to implement numerically. The experimentally measured 
flux noise is simply given by  the Fourier transform of $g_\ell ^V(t)$ i.e.
$S_{\Phi}(\omega)=\int dt~e^{i\omega t}   g_\ell ^V(t)$.
To analyze the scaling results for $S_{\Phi}$ we also need to calculate the
time-dependent equilibrium phase correlation function, defined as
$
g_{\theta}({\bf r},{\bf r'}, t, t')=<e^{i[\theta ({\bf r},t)-\theta({\bf r'},t')]}>.
$
In our explicit calculations we evaluate the zeroth-momentum correlation 
function, that is known to have only one dominant Lyapunov exponent 
\cite{ram}. Here we concentrate on the $T_{BKT}< T$ region, as was 
done in the Shaw {\em et al.} experiment. In the long distance  regime
\begin{equation} 
g_{\theta}(r)=Ar^{-\eta} e^{- r/{\xi(T)}},
\label{corrf}
\end{equation}
where the correlation length $\xi$ is expected to have the $BKT$ form, 
\begin{equation}
\xi(T)=\xi_0 e^{{b}/{\sqrt{T-T_{BKT}}}}, 
\label{corrl}
\end{equation}
with $T_{BKT}$, $\xi_0$ and $b$ constants determined in the calculations.
In Fig.\ref{1} we show the results for the equilibrium
time-averaged $g_{\theta}(r)$ for different temperatures, 
for both the TDGL (a) and RSJ (c) models in a $64\times 64$ 
array. In the insets we plot the corresponding $\xi(T)$ data
obtained from the correlation function calculations, together 
with their BKT-type fits. The fitting parameters are $\xi_0=0.230(1)$, 
$b=1.45(1)$ and $T_{BKT}=0.935(1)$ (inset in Fig.\ref{1}(a) TDGL)
 and $\xi_0=0.274(30)$, $b=1.41(10)$ and $T_{BKT}=0.917(10)$ (inset in 
Fig.\ref{1}(c) RSJ). These non-universal results compare well with the numbers 
obtained from a  more extensive Monte Carlo calculation that gave
 $\xi_0=0.205$, $b=1.6113$ and $T_{BKT}=0.9035$  \cite{ram}. 
The reasonable agreement between these and the MC results gives strong 
support to the reliability of our dynamic simulation algorithms \cite{eikmans}.

The results for  $g_{\ell}^V(t)$ are shown in Fig. \ref{1} (b) (TDGL) and 
\ref{1} (d) (RSJ), for an $\ell=32$ SQUID. As in the experiment we have 
restricted the analysis to the temperature regime with short 
correlation lengths in the range $\xi< \ell< L$. We tried different 
reasonable fitting functions for the data for $g_\ell^V(t)$ in the 
short and intermediate time regimes. From
stretched exponentials, that gave reasonable results, to a novel 
and somewhat comparable fit, with fewer parameters, that involves 
the modified Bessel function $K_\nu(t)$, with $\nu$ close to zero. 
The motivation for using the latter functions will become 
clearer below. First we show  the noise functions $S_\Phi(\omega)$ 
in Fig.\ref{2} (TDGL)(a) and (RSJ)(c). 
These figures were obtained from a direct fast Fourier transform 
of the $g_\ell^V(t)$ data. We  see there that the intermediate frequency
regime of $S_\Phi$ can be approximately fitted  by
a straight line leading to  $S_\Phi\sim 1/{\vert 
\omega\vert}^{\alpha(T)}$.  The $\alpha$ exponent has a 
weak non-monotonic temperature dependence but within a rather
narrow range, $0.85 \leq \alpha (TDGL)(T)\leq 0.95 $,
and $1.17\leq \alpha (RSJ)(T) \leq 1.27 $, 
as shown in the inset to Figs. \ref{2} (a) and (c). 
At lower frequencies there is a bending of $S_\Phi$ tending towards 
a flat behavior for the lower temperatures considered. We note that the TDGL 
$\alpha$ values are closer to the experimental values which were 
about one.  Although the RSJ results  have slightly larger 
$\alpha$'s  they can still be classified as anomalous 
$1/{\vert \omega\vert}$ noise. 
%%%%%%%%%%%%%%%%%%%%%%%%%%%%%%%%%%%%%%%%
%{figure 2}
%%%%%%%%%%%%%%%%%%%%%%%%%%%%%%%%%%%%%%%%
Based on these results we propose an analytic ansatz for 
$S_\Phi(\omega)$ of the form $S_{\Phi}(\omega)\sim \vert 
\Omega^2+\omega^2\vert^{-\alpha(T)/2}$. Here $\Omega$ is a 
characteristic frequency which in our calculations appears to be small.
Further justification for this ansatz will be given below. We can then 
obtain the corresponding expression for $g_\ell^V(t)$ from the inverse 
Fourier transform of the ansatz for $S_\Phi(\omega)$ giving,
\begin{equation}
g_\ell^V(t)=Ct^{\nu}\frac{\Gamma(1/2)}{\Gamma(\nu+1/2)}K_\nu(\Omega t),
\label{bessel}
\end{equation}
where $\nu\equiv (\alpha(T)-1)/2$, $\Gamma(x)$ is the Gamma function
and $C$ is a constant. We have used this result to fit 
the numerical data for $g_\ell^V$  and the fits are shown as 
continuous lines in Figs. \ref{1}(b) and \ref{1}(d). Note that based 
on the calculated values for $\alpha$, the Bessel function index 
$\nu$ takes the values, $-0.075\leq \nu (TDGL)\leq -0.025$ and 
$0.085\leq \nu (RSJ)\leq 0.135$ (recall that $K_{-\nu}(x)=K_{\nu}(x)$). 
It is clear that the TDGL model is rather close to the $K_0$ result, 
while the RSJ is slightly different. To further understand our 
numerical results, given that $\nu$ is rather close 
to zero, it is instructive to first consider the case 
when $\alpha=1$, which leads to $g_\ell^V\propto K_0(\Omega t)\propto 
e^{-\Omega t}\int_0^{\infty}e^{-2\Omega \tau}
{d\tau}[{{\tau({\tau+t})}]^{-1/2}}$,
with Fourier transform $\propto (\Omega^2+\vert \omega\vert^2)^{-\frac{1}{2}}$. 
In the limit when $\omega\gg \Omega$, 
$S_\Phi(\omega)\sim 1/{\vert \omega\vert}$, 
while in the opposite limit it is a constant and the noise is 
white, just as in the experiment. The large $\omega$ regime corresponds 
to the logarithmic short time region of $K_0(\Omega t)$, whereas the lower
$\omega$ regime corresponds  to the exponential one.  
We can go further by looking at the definition of 
$g_\ell^V$ given in Eq.(\ref{gvp}), which implies a coarse
grained $N(t)\sim t^{-1/2}e^{-\Omega t}$
(for $\alpha=1$). Here we are making use of the  ergodic theorem.
 We can extend the analysis given above to $\nu\neq 0$, which yields
$N(t,\nu (T))\sim t^{-(1/2-\nu)}e^{-\Omega t}$. The evaluation of the
time Fourier transform of $<N (\tau,\nu (T))N (\tau+t,\nu (T))>$ 
gets us back to our ansatz for $S_\Phi(\omega)$.
This is an interesting result that can be interpreted in terms of 
an incoherent superposition of a series of independent random wave front
events produced by the entering and leaving of vortices from 
the effective SQUID area. The coarse grained events can be assumed to be 
Poisson distributed in time so as to produce  a random pulse train with 
the explicit time dependence for $g^V$ given above (see Ref. \cite{schonfeld}).

Another important experimental result was the test of 
the dynamic scaling predictions in frequency space. 
We have carried out an equivalent scaling analysis in time.
%, for our data is more extensive in this domain.
The general dynamic scaling hypothesis assumes that 
$g_\ell^V=\xi^{\beta}F_\ell (t/{\tau_\xi},L/{\xi},\ell/{\xi})$.
We find that, as in the experiment, for $\xi< \ell < L$,
we can collapse the data into a single 
curve by considering only the time dependent part of the 
scaling function, i.e. $g_\ell^V\propto \xi^{\beta}G(t/{\tau_\xi})$,
with $\beta$ an exponent and the relaxation time $\tau_\xi\propto\xi^z$, 
thereby defining the dynamical critical exponent, $z$.
We simulated a $64\times 64$ array for different temperatures and 
we were able to  collapse all the data into one single curve 
for different $T$'s and $\ell$'s using this criterion.
The data collapsed corresponds to roughly a
decade in $\tau_\xi$.  We used data for $7$ different
$T$'s with $\ell=16$ and $\ell=32$, for $\xi (T)\epsilon [2.2, 8]$.
The accessible length scales and time scales 
are limited by the available computer power. For a given 
$\xi$, we used an $L>8\xi $ lattice and simulated for 
$100$-$1000 \tau_{\xi}$. The scaling results are shown in 
Fig. \ref{2}(b) (TDGL) and  \ref{2}(d) (RSJ). From the scaling 
analysis we determined $z$ and obtained the values of $z(TDGL)\sim 2$ and 
$z(RSJ)\sim 0.9$, as shown in the insets to Figs. \ref{2}(b) 
and \ref{2}(d). The estimated $z(TDGL)\sim 2$  exponent is consistent with the 
value  obtained in the experiment \cite{shaw}. 

We have also made animations of the vortex motions above $T_{BKT}$. There we
see that the dynamics is dominated by vortex clusters formed above
$T_{BKT}$.  One can thus link the anomalous values of the 
$\alpha$ exponents to the anomalous way in which single vortices 
diffuse through the clusters.

In conclusion, we have studied the flux noise in the TDGL
and RSJ models with the goal of providing an understanding
to the recent interesting experiments by Shaw {\em et al.} \cite{shaw}. 
We find that by following the prescriptions defined in the experiment,
both models lead to anomalous vortex dynamics, with  the
TDGL results being in closer agreement to the experimental
ones. It is possible that the actual samples that are made of large area
Nb islands deposited on a copper substrate have a very small resistance
to ground and thus the TDGL component may dominate the bond or RSJ
term \cite{shaw2}. We also have studied the effect of a small 
magnetic field and we find that the results change significantly. 
The $\alpha$ exponent acquires a stronger temperature
dependence.  Furthermore, we have preliminary results for the 
$T<T_{BKT}$ regime and the $\alpha$-values obtained are also  
indicative of anomalous vortex dynamics. 
We will report on these and other results in a separate publication.

\vskip 0.3cm

We thank T. Shaw for a copy of Ref. \cite{shaw} prior to publication, 
for careful reading of this paper and for several useful informative 
exchanges. This work was supported in part by the Dutch organization 
for fundamental research (FOM), by NSF grant No. DMR-9521845 (JVJ) and by
the Bavarian ``FORSUPRA" program (TJH).

\newpage

%%%%%%%%%%%%%%%%%%%%%%%%%%%%%%%%%%%%%%%%%%%%%%%%%%%%%%%%%%%%%%%%%%%%%%%
%
%                         Figures:
%
%%%%%%%%%%%%%%%%%%%%%%%%%%%%%%%%%%%%%%%%%%%%%%%%%%%%%%%%%%%%%%%%%%%%%%%
%
\begin{figure}[h]
\caption{Results from simulations of a $64\times 64$ array with PBC.
 Phase correlation functions $g_{\theta}(x)$ obtained from TDGL(a)
and RSJ(c) simulations. The different curves correspond to:
$T=1.08$ ($\circ$), $T=1.12$ (squares), $T=1.24$ (diamonds)
and $T=1.32$ ($\bigtriangleup$). The continuous lines are fits
to Eq. (\ref{corrf}).
Insets to (a) and (c) give the $\xi (T)$ results for TDGL and RSJ
dynamics, with the continuous lines fits to the BKT form given in
Eq.(\ref{corrl}), respectively. In (b) and (d) we plot
$g_{\ell}^V(t)$ for $\ell=32$. The temperatures and
notation is the same as in (a) and (c). The solid lines are fits to
Eq.(\ref{bessel}).}
\label{1}
\end{figure}
\begin{figure}[h]
\caption{Flux-noise $S_\Phi(\omega)$ vs $\omega$  plotted on
log-log scale for TDGL(a) and RSJ(c) dynamics for the same parameters
as in Fig. \ref{1}. The different temperature curves have been
shifted upwards by a constant distance for clarity. From top to
bottom (a) $T=1.08$, $1.16$, $1.24$, $1.32$, $1.40$ and $1.48$  and
(c) $T=1.08$, $1.10$, $1.12$, $1.20$,$1.28$ and $1.36$. In the insets
to panels (a) and (c) we plot the exponent $\alpha(T)$ obtained from
a least squares fit to the quasi-linear frequency regime.
The scaling plots for $g_l^V(t)$ are shown in panels TDGL(b)
($T=1.12$, $1.16$, $1.20$, $1.24$, and $1.28$),
and RSJ(d) ($T= 1.12$, $1.16$, $1.20$, $1.24$, $1.28$, $1.32$, and
$1.36$), with  $\ell=16$ ($+$)  and  $\ell=32$ ($\circ$).
In the insets to panels (b) and (d) we show the scaling of
$\tau_\xi$ vs  $\xi$ on a log-log plot, for $\ell=32$ ($\circ$) and
$\ell=16$ (square). The continuous (dot-dashed) 
line is a linear fit to the
$\ell=32$ ($\ell=16$) results.}
\label{2}
\end{figure}

\begin{references}
\bibitem{cap} These arrays have zero capacitances thus zero quantum 
fluctuations.
\end{references}
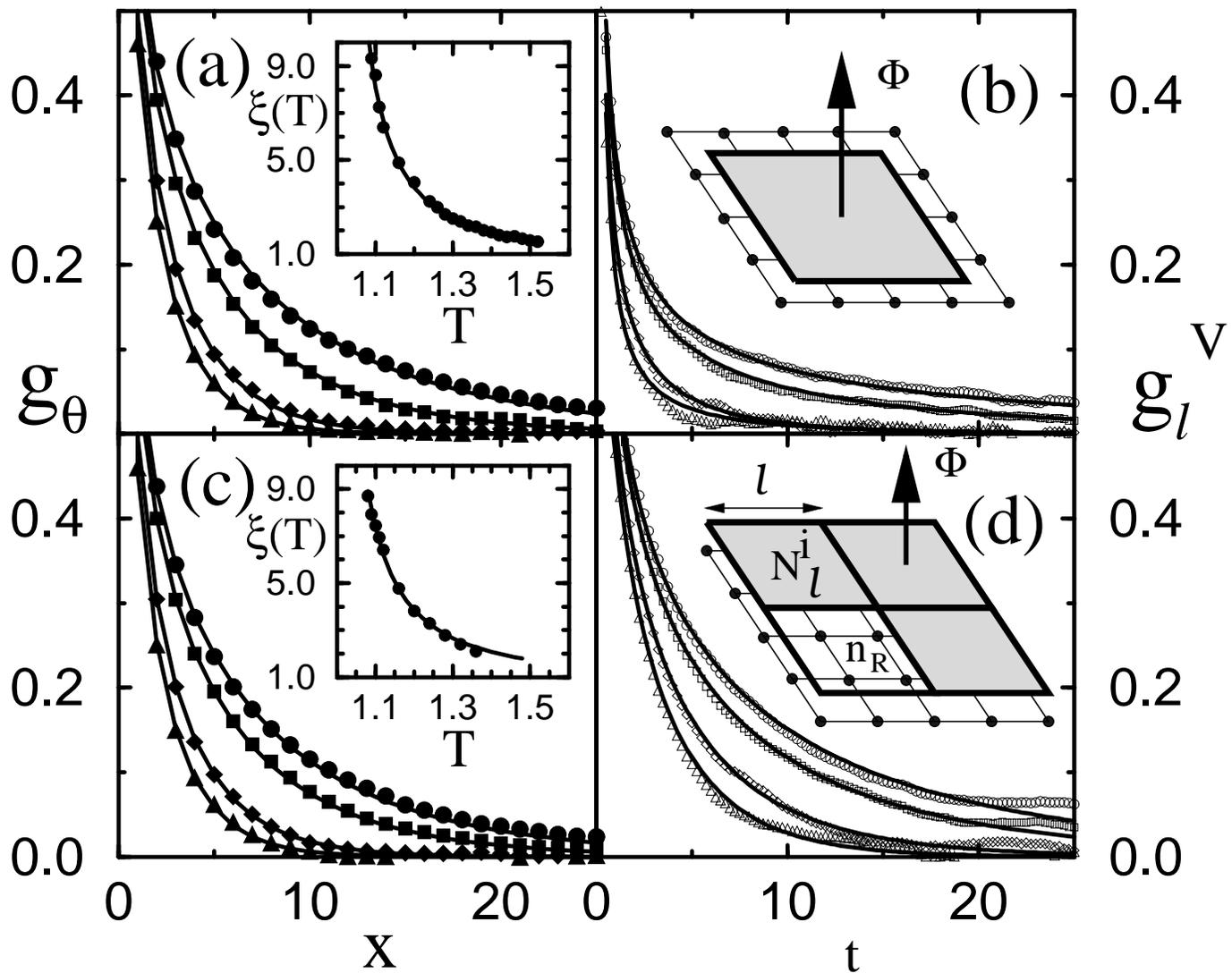
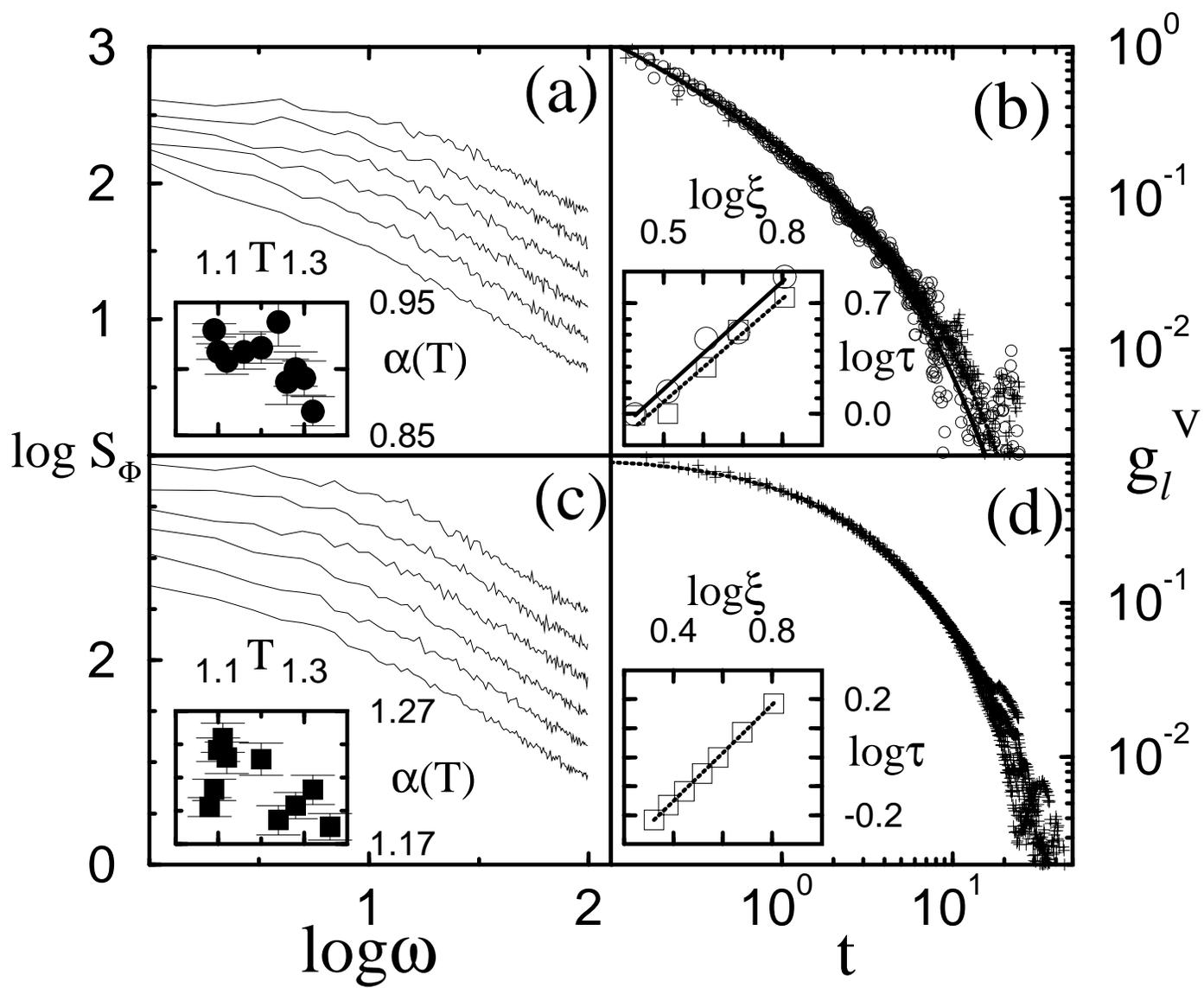
\end{document}